\documentclass[prl,twocolumn,preprintnumbers,amsmath,amssymb,showpacs,nofootinbib,floatfix]{revtex4}

\usepackage{graphicx,bm}

\makeatletter
\def\graphicscale{\twocolumn@sw{0.3}{0.4}}
\def\graphicthreescale{\twocolumn@sw{0.3}{0.4}}

\begin{document}

\title{Off-equilibrium finite-size method for critical behavior analyses}

\author{Matteo Lulli,$^{1,2}$ Giorgio Parisi,$^{3,4}$ and
Andrea Pelissetto$^{3,4}$}

\address{$^1$ Physics of Fluids Group, University of Twente, 7500AE Enschede,
   The Netherlands}

\address{$^2$ Department of Applied Physics, Technische Universiteit Eindhoven,
    5600 MB Eindhoven, The Netherlands}

\address{$^3$ Dipartimento di Fisica di ``Sapienza," Universit\`a di
    Roma,I-00185 Roma, Italy}

\address{$^4$ INFN, Sezione di Roma I, I-00185 Roma, Italy}

\date{\today}

\begin{abstract}
We present a new dynamic off-equilibrium method for the study of 
continuous transitions, which represents a dynamic generalization of the 
usual equilibrium cumulant method. Its main advantage is that 
critical parameters are derived from numerical data obtained much before 
equilibrium has been attained. Therefore, the method is particularly 
useful for systems with long equilibration times, like spin glasses. 
We apply it to the three-dimensional Ising 
spin-glass model, obtaining accurate estimates of the critical exponents and 
of the critical temperature with a limited computational effort.
\end{abstract}

\pacs{64.60.Ht,64.60.F-,05.70.Jk,75.10.Nr,75.10.Hk}
%% 05.10.Cc	Renormalization group methods
%% 05.70.Fh	Phase transitions: general studies
%% 05.70.Jk	Critical point phenomena 
%% 64.60.an	Finite-size systems
%% 64.60.De	Statistical mechanics of model systems (Ising model, Potts model, field-theory
%% models, Monte Carlo techniques, etc.)
%% 64.60.F-	Equilibrium properties near critical points, critical exponents
%% 64.60.Ht	Dynamic critical phenomena
%% 75.10.Hk	Classical spin models
%% 75.10.Nr	Spin-glass and other random models
%% 75.40.Cx	Static properties (order parameter, static susceptibility, heat capacities,
%% critical exponents, etc.)
%% 75.50.Lk	Spin glasses and other random magnets

\maketitle

% ========================= BODY =========================

Monte Carlo simulations combined with finite-size scaling (FSS) methods
are, at present, the most successful tool for the identification of 
continuous transitions and the determination of the critical parameters. 
In this approach there are two main obstacles to a precise determination 
of the critical parameters. On one side, scaling corrections, related to the 
subleading irrelevant renormalization-group (RG) operators, may mask the 
asymptotic critical behavior, which shows up only when the system size $L$ 
becomes large. On the other side, the Monte Carlo dynamics 
becomes increasingly slow as the critical point is approached, so that 
thermalization and equilibrium autocorrelation times become large as $L$ 
increases, hampering large-$L$ simulations. These problems are particularly
serious in systems with quenched disorder. They occur, for instance, 
when studying the paramagnetic-glassy transition in the $\pm J$ Ising model 
with random ferromagnetic and antiferromagnetic couplings, which 
represents a standard theoretical laboratory to understand the effects of 
quenched disorder and frustration. Its Hamiltonian
is given by \cite{EA-75}
\begin{equation}
H = - \sum_{\langle xy\rangle} J_{xy} \sigma_x \sigma_y,
\end{equation}
where the sum runs over all nearest neighbors $\langle xy\rangle$ 
of a cubic lattice, 
$\sigma_x = \pm 1$ are Ising spins, and 
$J_{xy}$ are quenched random couplings that take the values $\pm 1$ with 
equal probability. At the transition and, even worse, 
in the low-temperature phase, the standard Metropolis dynamics is very slow,
so that equilibration becomes very difficult, even for relatively small 
systems. Moreover, equilibration times depend strongly on the disorder
realization,
so that a sample-by-sample analysis is needed to guarantee that all 
measurements are obtained from well-equilibrated samples \cite{Janus-10}.
 At present, even by
using dedicated machines \cite{Janus-13}, it seems impossible to go much beyond 
sizes $L=30$-40. 

In this work we discuss a dynamic method to determine the critical 
temperature and the critical exponents. We will discuss it in the context of
the $\pm J$ Ising model, but the method and the results are completely general,
so it can be applied to the study of any continuous transition in pure 
or disordered systems. The method, which does not
require the system to be in equilibrium, has two advantages. First,
the difficult and time consuming (at least in disordered
systems) task of verifying equilibration
is no longer needed. Second, 
we can stop the simulation much before equilibrium has been reached, 
saving a considerable amount of computing time. The method we discuss
is somewhat different from previous off-equilibrium methods 
(see, e.g., 
Refs.~\onlinecite{OI-01,NEY-03,PC-05,PGRLT-06,Janus-z,Roma-10,Nakamura-10,%
FM-15,LPSY-15} 
and references therein).
Indeed, in most
of those works it is generally assumed that $L$ is so large that finite-size
effects are negligible, a condition that is easily satisfied in pure systems
but not in the random case. On the contrary, we will use the finite-size
dependence of physical observables to estimate the critical parameters. 
Our method is essentially an off-equilibrium generalization of the 
usual Binder crossing method.

As in FSS equilibrium computations, we begin by considering a 
RG invariant quantity $U(t,L,\beta)$ as a function of the 
Monte Carlo time $t$, inverse temperature $\beta$, and system size $L$.
According to RG, for $L$ and $t$ large and close to the critical point $\beta_c$,
$U(t,L,\beta)$ scales as \cite{HH-77,Suzuki-76,Suzuki-77,JSS-89,CG-05,OI-07}
\begin{equation}
U(t,L,\beta) = f_R(t L^{-z},\epsilon) + u_\omega(\beta) L^{-\omega} 
               g_R(t L^{-z},\epsilon) + \ldots
\label{eq:R_vs_tLmz}
\end{equation}
where next-to-leading scaling corrections have been neglected.
Here $\omega$ is the leading correction-to-scaling exponent,
$u_\omega(\beta)$ the associated nonlinear scaling field satisfying 
$u_\omega(\beta_c)\not=0$, $\epsilon = u_\beta(\beta) L^{1/\nu}$,
where $u_\beta\approx \beta - \beta_c$ is the temperature nonlinear
scaling field which parametrizes the distance from the critical point. 
Equation~(\ref{eq:R_vs_tLmz}) depends also on the dynamic critical exponent
$z$, which represents an additional parameter to be determined. To avoid 
any reference to $z$, we now reparametrize the time evolution in terms of the 
correlation length $\xi$, or, better, in terms of the RG invariant ratio
$R_\xi = \xi/L$. When using an initial disordered configuration, $\xi$ 
is an increasing function of $t$, any function of $t L^{-z}$ 
can be equivalently reexpressed in terms of $R_\xi$, so that we write
\begin{equation}
U(t,L,\beta) = \hat{f}_U(R_\xi,\epsilon) + u_\omega(\beta) L^{-\omega} 
               \hat{g}_U(R_\xi,\epsilon) + \ldots,
\label{eq:R_vs_Rxi}
\end{equation}
which is defined for $R_\xi \le R_{\xi,\rm eq}(\epsilon)$, where 
$R_{\xi,\rm eq}(\epsilon)$ is the equilibrium value of $R_\xi$ for the 
given $\epsilon$. Equation~(\ref{eq:R_vs_Rxi}) is the basic relation we use
to compute critical temperature and exponents. Indeed, ignoring scaling 
corrections, close to the critical point Eq.~(\ref{eq:R_vs_Rxi}) can 
be expanded in $\epsilon$, obtaining
\begin{equation}
U(t,L,\beta) = \hat{f}_U(R_\xi,0) + (\beta - \beta_c) L^{1/\nu} 
               \hat{f}_U'(R_\xi,0) + \ldots
\label{eq:R_vs_Rxi-expan}
\end{equation}
At fixed $R_\xi$, the quantity $U(t,L,\beta)$ behaves exactly as in the 
equilibrium case: $\beta_c$ is determined as the crossing point 
and $\nu$ is obtained by computing the slope at $\beta_c$. 
However, in this formulation
equilibration is not needed. Equation~(\ref{eq:R_vs_Rxi}) is valid for any
value of $R_\xi$, so one might think of choosing a small value for such 
a parameter, reducing significantly the length of the runs. However, one 
must not forget that the method is intrinsically a finite-size method; hence,
it can only work if finite-size effects are not too tiny, and this, in turn,
requires $R_\xi$ to be not too small. Mathematically, these considerations 
can be understood by considering Eq.~(\ref{eq:R_vs_Rxi-expan}). 
The method is expected to be precise if the coefficient $\hat{f}_U'(R_\xi,0)$ 
is not too small. Such a coefficient depends on $R_\xi$ and it is expected 
to increase with $R_\xi$. In particular, it is expected to be very small 
for $R_\xi$ small, so that, if one chooses a small value of $R_\xi$, 
the crossing becomes undetectable, unless statistical errors are very tiny. 
Hence, $R_\xi$ should be chosen small, but still large enough to have 
a reasonable sensitivity of the results on system sizes. 

To validate the method, we apply it to the determination of the critical point 
and of the critical exponents in the $\pm J$ Ising model. 
We perform large-scale
simulations
on cubic lattices of volume $L^3$, with $8\le L\le 64$, considering five 
values of $\beta$ between 0.880 and 0.910. Statistics is a crucial factor 
in the analysis and hence we consider a very large number $N_s$ of samples 
for each $L$ and $\beta$. Typically, $N_s$ varies between 
$3\cdot 10^5$ and a few million.
Only for $L=48,64$ is $N_s$ smaller: $N_s = 6\cdot 10^4, 10^4$ in these cases.
Essentially all runs end when the system is still 
out of equilibrium. In most of the cases, data extend only up $R_\xi \approx
0.5$, in some cases even less (at equilibrium \cite{Janus-13} 
$R_\xi = 0.652(3)$ at the critical point). 
Simulations were performed on a small GPU cluster using a very efficient
asynchronous multispin coding technique \cite{BL-14}.
In each run we simulate together $32k$ different disorder 
realizations with four replicas for each disorder realization. The value of 
$k$ is tuned for each $L$ to have the best performance of the GPUs. 
As a result, one spin flip takes 2.9 ps (essentially for all sizes)
on the 
GTX Titan, the fastest GPU we have. The simulations presented here 
took approximately 3.1 CPU years of the GTX Titan GPU. 

To apply Eq.~(\ref{eq:R_vs_Rxi}) we must define the quantities $U$. We 
consider 5 different Binder cumulants defined in terms of the overlap between
two different replicas $q_x = \sigma^{(1)}_x\sigma^{(2)}_x$. 
Moreover, we must somehow 
parametrize the scaling functions $\hat{f}_U(R_\xi,\epsilon)$ and 
$\hat{g}_U(R_\xi,\epsilon)$. Since the data belong to a small temperature 
interval, we use the expansion (\ref{eq:R_vs_Rxi-expan}) to first order in
$\epsilon$. We have also performed some analyses
using a second-order approximation, without observing significant differences.
As for the correction-to-scaling function, we have verified that 
we can assume it to be independent of temperature. Finally, we should make 
approximations for the nonlinear scaling fields. Relying on the analysis 
of Refs.~\onlinecite{HPV-08-lett,HPV-08}, 
we set $u_\beta(\beta) = \beta - \beta_c$ and $u_\omega(\beta) =
1$, neglecting the additional corrections. Given the small temperature
interval we consider, these approximations should hold quite 
precisely. Hence, each $U(t,L,\beta)$ was fitted to 
\begin{equation}
U(t,L,\beta) = P_1(R_\xi) + P_2(R_\xi) (\beta - \beta_c) L^{1/\nu} + 
               P_3(R_\xi) L^{-\omega},
\end{equation}
with $P_1(R_\xi)$, $P_1(R_\xi)$, and $P_3(R_\xi)$ polynomials of 
degree 6, 3, and 3, respectively. The fit of the five renormalized couplings
is quite complex, as we take $\omega$, $\beta_c$, $\nu$, and the coefficients
of the polynomials as free parameters. As a whole, there are 78 free parameters
that must be optimized. In the fits we have not taken into account 
the time correlations among data at the same $\beta$ and $L$, so that 
statistical errors (computed using the jackknife method)
are not {\em a priori} optimal. To verify that such neglect is not relevant 
for the final estimates, we have performed some fits of a single cumulant 
taking time correlations into account. The corresponding estimates and 
error bars are essentially equal to those obtained without including 
statistical correlations.

As usual in this type of analyses, the most difficult issue is the 
estimation of the systematic errors due to the neglected correction terms. 
This is very important here, since the attainable values of $L$ are quite
small. We have thus performed fits with several types of cuts. We perform fits 
including each time only data satisfying $L\ge L_{\rm min}$, $\xi\ge \xi_{\rm
min}$, and $R_\xi\ge R_{\xi,\rm min}$, considering several values for 
$L_{\rm min}$, $\xi_{\rm min}$, and $R_{\xi,\rm min}$. Results obtained 
by taking $3 \le \xi_{\rm min} \le 5$, $8\le L_{\rm min} \le 12$, and 
$0 \le R_{\xi,\rm min} \le 0.4$ show some 
scatter, which is somewhat larger than statistical errors, indicating that 
the neglected systematic effects may be as important as the statistical ones.
The most crucial parameter is $\xi_{\rm min}$. When such a parameter
is increased from 3 to 4, the exponent $\omega$ decreases sharply, by more 
than one error bar, while $\beta_c$ increases. Such a systematic
drift occurs also when $\xi_{\rm min}$ is further increased to 5, but now
the change is much less than one error bar. Therefore, the results we quote 
correspond to fits with $\xi_{\rm min} = 4$. 
For such a value of $\xi_{\rm min}$ we obtain
$\beta_c = 0.911(2)$, 0.916(4), 0.909(4) for $L_{\rm min} = 8,10,12$ and 
$R_{\xi,\rm min} = 0$, and 
$\beta_c = 0.911(2)$, 0.909(2), 0.909(3) for 
$R_{\xi,\rm min} = 0$, 0.2, 0.4 and $L_{\rm min} = 8$.
No systematic trends can be observed, all estimates being consistent 
within errors. Except for one estimate, all results (with their errors)
we are quoting here
lie in the interval $0.906\le \beta_c \le 0.913$. Therefore, we take
$
\beta_c = 0.910(4)
$
as our final estimate. The error, which is twice the error affecting 
the results with $L_{\rm min} = 8$, is somewhat subjective and should take
into account the effect of the neglected next-to-leading scaling corrections. 
Analogously, we can estimate $\omega$ and 
$\nu$ obtaining
\begin{equation}
\omega = 1.3(2), \qquad \nu = 2.47(10).
\end{equation}
The estimates of $\omega$ are strongly 
correlated with those of $\beta_c$: the larger $\beta_c$, the smaller $\omega$
is.  If $\beta_c = 0.906$, fits keeping $\beta_c$ fixed give $\omega\approx 1.5$,
while $\omega\approx 1.1$ is obtained by fixing $\beta_c = 0.914$.
The exponent $\nu$ is instead much less 
correlated with $\beta_c$, changing at most by $\pm 0.03$ when $\beta_c$ varies 
by $\pm 0.004$. 

\begin{figure}[tb!]
\begin{center}
%%\begin{tabular}{c}
%% \includegraphics[width=6cm,angle=-90]{cutRxScaledTeX.ps} \\
 \includegraphics[scale=0.6]{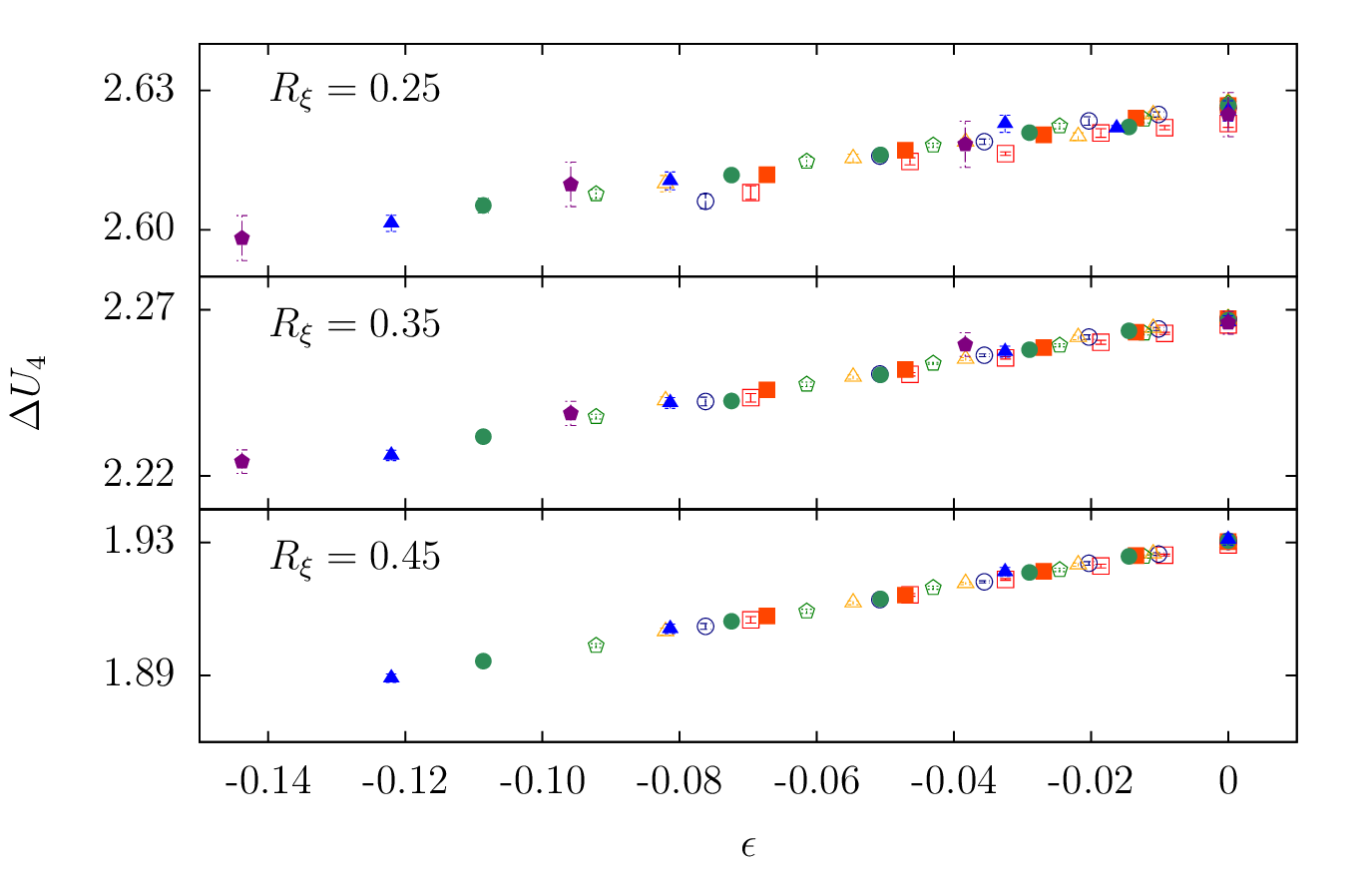} \\
%%\end{tabular}
\end{center}
\caption{Plot of $\Delta U_4$ versus $\epsilon = (\beta - \beta_c) L^{1/\nu}$
for $R_\xi = 0.25$ (top), 
$R_\xi = 0.35$ (middle) and $R_\xi = 0.45$ (bottom). We set $\beta_c = 0.910$,
$\omega = 1.3$ and $\nu = 2.47$. Here $U_4$ is the standard Binder cumulant,
as defined in Refs.~\onlinecite{HPV-08,Janus-13}.
Symbols: empty square ($L=8$), empty circles ($L=10$), empty triangles ($L=12$), empty pentagons ($L=16$), filled squares ($L=20$), filled circles ($L=24$), filled triangle ($L=32$), filled pentagon ($L=48$).%stars ($L=8$), empty squares ($L=12$), filled squares ($L=16$), empty circles ($L=24$), filled circles ($L=32$).
}
\label{fig:U4}
\end{figure}

To show the quality of the results, in Fig.~\ref{fig:U4}, we report 
$\Delta U_4$, defined by 
\begin{equation}
\Delta U_4(\beta,L,R_\xi) = U_4(\beta,L,R_\xi) - P_3(R_\xi) L^{-\omega},
\end{equation}
versus $\epsilon$. 
We consider $R_\xi = 0.25$, 0.35, and 0.45. Very good
scaling is observed, confirming the correctess of the scaling Ansatz and 
the accuracy of the estimates of the critical exponents. Note also that 
the data lie on an essentially straight line, validating our choice 
of expanding $f_U(R_\xi,\epsilon)$ to first order in $\epsilon$. From the 
figure, we can also clarify why a large number of samples, of order $10^6$, 
is needed to estimate the critical parameters. 
For instance, $U_4$ at $R_\xi = 0.35$ varies by 0.04 within our temperature
interval.  Therefore, the temperature dependence
of the data can be observed only if the errors on $U_4$ 
are significantly less than
$10^{-2}$, for instance, if they are equal to $10^{-3}$. 
Since errors scale as $a/\sqrt{N_s}$ with $a\approx 1$ for all 
values of $L$, a $10^{-3}$ error is obtained by taking
$N_s\approx 10^6$. Note that this requirement is not specific of the 
off-equilibrium method we use. 
Also equilibrium analyses require $N_s$ to be large
\cite{HPV-08-lett,HPV-08,Janus-13}.

\begin{table*}[!t]
\caption{Estimates of $T_c$ and of the critical exponents by off-equilibrium
methods. Results of Refs.~\onlinecite{HPV-08,Janus-13} are obtained 
in equilibrium simulations. The exponents $\beta$ and $\gamma$ are 
related to $\nu$ and $\eta$ by $\beta = \nu (1+\eta)/2$, $\gamma = \nu (2 -
\eta)$. }
\label{estimates}
\begin{center}
\begin{tabular}{ccccccccc}
\hline\hline
 & $T_c$  & $\nu$  & $\eta$   & $\omega$ & $z$  & $\beta$ & $\gamma$ & $z\nu$ \\
\hline
%% Ogielski \cite{Ogielski1985}
%% & $1.175(25)$ &            &              &          & 
%% $6.1(3)$   &           &          &          
%% \hline
%% Campbell \cite{BernardiPakashCampbell1996}
% & $1.17(1)$   &            & $-0.25(2)$   &          
%% & $6.0$      &           &           \\
%%         \cite{MariCampbell1999}          & $1.20(1)$   &            
%% & $-0.21(2)$   &          &            &           &        \\
%%         \cite{MariCampbell2002}          & $1.195(15)$ &            &   
%% & $2.9(6)$ & $5.65(15)$ &           &        \\
Ref.~\onlinecite{OI-01}    
& $1.12(12)$  &    &   &   &   &    &   & $5.7(5)$\\
Ref.~\onlinecite{NEY-03}     
& $1.17(4)$ & $1.5(3)$ &   &   & $6.2(2)$  &  & $3.6(6)$  &\\
Ref.~\onlinecite{PC-05}     
& $1.19(1)$ &   & $-0.22(2)$ &  & $5.7(2)$  &       &      &  \\
Ref.~\onlinecite{PGRLT-06} & 
$1.154(3)$ &  &    &   &   & $0.52(9)$ &   & \\
Ref.~\onlinecite{Janus-z} 
  & & & & & 6.86(16) & & &\\
Ref.~\onlinecite{Roma-10} 
& $1.135(5)$  &    &   &   &   &    &   & \\
Ref.~\onlinecite{Nakamura-10}  
& $1.18(1)$ & $1.40(5)$ & $-0.20(1)$ &  &   &   &   &\\
Ref.~\onlinecite{FM-15,private} 
& 1.115(15) & $2.2(3)$ & $-0.380(7)$ &  & 6.79(6) &
      &   &   \\
Ref.~\onlinecite{LPSY-15} 
  & & & & & 5.85(9) & & &\\
this work 
& $1.099(5)$  & 2.47(10) & $-0.39(1)$ & $1.3(2)$ & 6.80(15) & & &\\
\hline
Ref.~\onlinecite{HPV-08}   
  & $1.109(10)$ & $2.45(15)$ & $-0.375(10)$ & $1.0(1)$ &  &   &  & \\
Ref.~\onlinecite{Janus-13}   
& $1.1019(29)$& $2.562(42)$& $-0.3900(36)$& $1.12(10)$&   &   &   &\\
\hline\hline
\end{tabular}
\end{center}
\end{table*}

It is interesting to compare these results with previous ones,
see Table~\ref{estimates} (older estimates are summarized 
in Ref.~\onlinecite{KKY-06}).  For the 
critical-point position, our estimate $T_c = 1/\beta_c = 1.099(5)$ agrees 
within errors with the 
estimates $T_c = 1.102(3)$ and $T_c = 1.109(10)$ of
Refs.~\onlinecite{Janus-13,HPV-08}, obtained from the analysis of equilibrium 
results.
Our error is larger than that reported in Ref.~\onlinecite{Janus-13}, 
but note that our final error includes a subjective estimate of the 
systematic error. Had we reported only the statistical error
for $L_{\rm min} = 8$, we would have obtained the same accuracy. 
The estimates of $\nu$ are also consistent, while our 
final estimate of $\omega$ 
is slightly larger, though still consistent within error 
bars, than previous ones. 
The off-equilibrium estimates of Ref.~\onlinecite{FM-15} are consistent 
with ours, but less precise.
Previous dynamic 
estimates of $T_c$ are instead not consistent.
It is now clear that the reported 
errors are underestimated, as a consequence of the neglect of 
the subleading scaling corrections in the analyses. 

The method we have discussed represents
a significant improvement with respect to equilibrium analyses. 
Indeed, since the scaling variable is $t L^{-z}$, the time needed to extend 
Metropolis 
runs from any value of $R_\xi$ to equilibrium scales as $L^z$, i.e., as $L^7$
given that $z \approx 7$ for the Ising spin glass. Therefore, the advantage 
is very large and increases rapidly with $L$. To make a fair comparison 
with equilibrium studies, we should, however, take into account that in those
studies one combines the parallel-tempering method \cite{Parallel-tempering}
with the Metropolis or heat-bath algorithm.
It is not clear how equilibration times scale for this 
combined algorithm, and in particular, how long it takes to thermalize 
the hard samples. However, the results reported in 
Ref.~\onlinecite{Janus-13} are consistent with 
a sample-dependent time that scales as $L^2$ for the samples that equilibrate
fast and as $L^7$ for those that are slower. 
The off-equilibrium method is still significantly faster.
A more direct comparison can be obtained by 
using the results published in Ref.~\onlinecite{Janus-13}. 
In our simulations at the critical point, runs extending up to 
$R_\xi \approx 0.5$ require $2.5\cdot 10^6$, $16\cdot 10^6$ Metropolis sweeps
for $L=24$ and 32, respectively. In the parallel-tempering simulations 
for $L=32$ of Ref.~\onlinecite{Janus-13}, 
the number of iterations discarded for thermalization 
varies between $8\cdot 10^6$ and $500\cdot 10^6$ (on average $13\cdot 10^6$)
sweeps. Taking into account that 22 systems at different temperature
are simulated together, our simulations are shorter
by a factor of 10 at least. If one were stopping the off-equilibrium 
runs at $R_\xi = 0.40$, one would 
gain an additional factor of 3 for this value of $L$. 

In spite of the significant improvement with respect to equilibrium
studies, the computing time needed for a simulation scales as $L^z$
even off-equilibrium, since we need to collect data at fixed $R_\xi$ for 
all values of $L$. 
This requirement makes our method not suitable
to investigate large system sizes. Since errors are independent of system size
as a consequence of the absence of self-averaging,
the Monte Carlo time needed to obtain the same statistical 
errors scales also as $L^z$.
This explains why we have not considered lattices with $L > 64$.
If we increase $L$, we should increase $t$ at the same time, making simulations
far too long. 

\begin{figure}[tb!]
\begin{center}
\begin{tabular}{c}
 \includegraphics[scale=0.6]{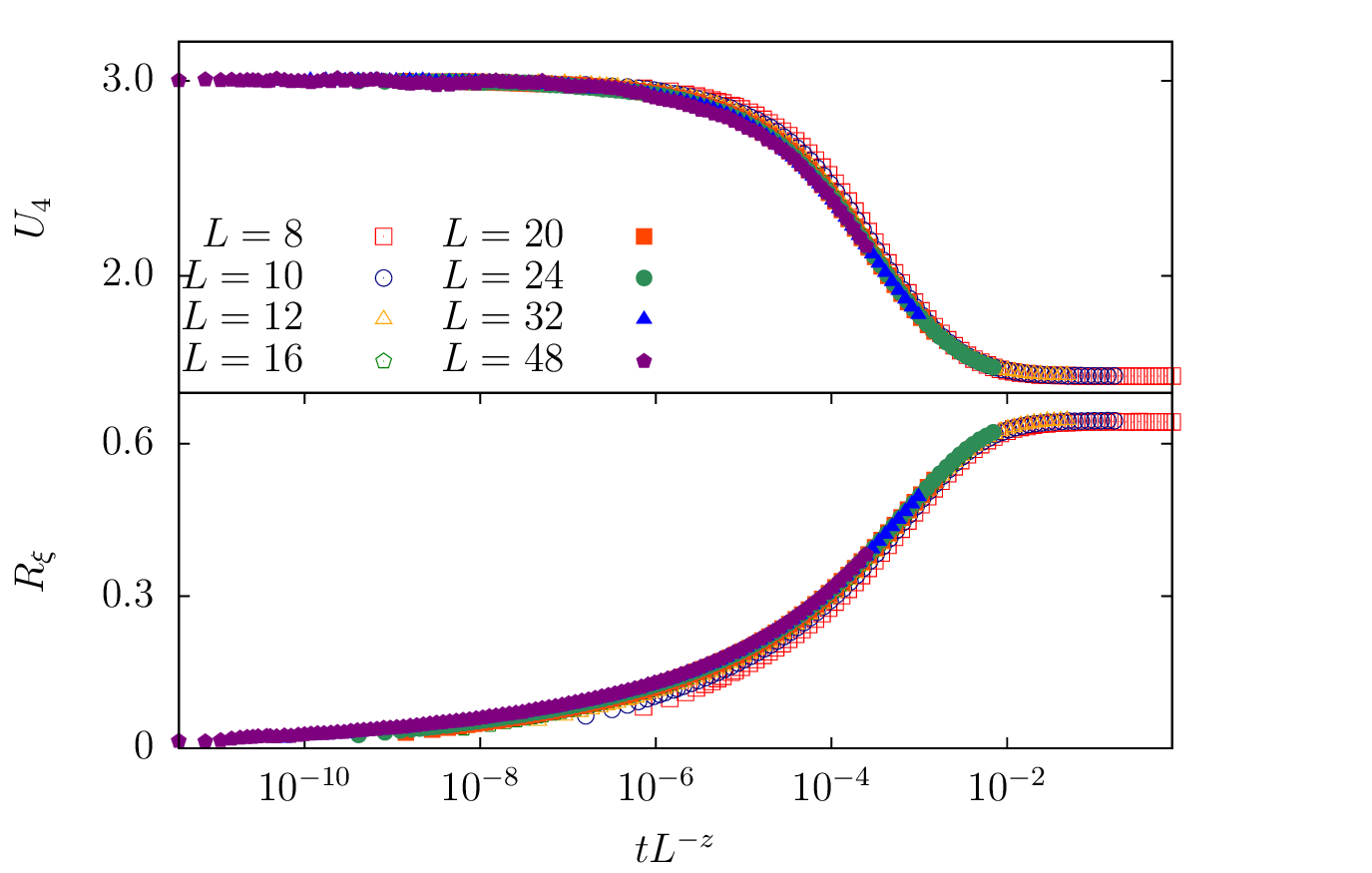} \\
\end{tabular}
\end{center}
\caption{Plot of $U_4$ and $R_\xi$ versus $t L^{-z}$ for $\beta = 0.910$. 
We set $z = 6.80$.
}
\label{fig:U4RxivstL}
\end{figure}

The analysis we have performed for the renormalized couplings can be extended 
to the susceptibility. The finite-time scaling behavior can now be written
as 
\begin{eqnarray}
\ln \chi &=& (2 - \eta) \ln L + P_1(R_\xi) + 
       (\beta -\beta_c) L^{1/\nu} P_2(R_\xi) + \nonumber \\
&& \quad L^{-\omega} P_3(R_\xi) + P_4(\beta).
\end{eqnarray}
where the last term $P_4(\beta)$ is the contribution of the nonlinear 
scaling field associated with the magnetic field, see 
Ref.~\onlinecite{HPV-08-lett,HPV-08} for a discussion. 
A good parametrization is obtained 
by taking $P_1(R_\xi)$, $P_2(R_\xi)$, $P_3(R_\xi)$ as polynomials of degree
6, 3, 3, respectively, as before. For the $P_4(\beta)$, we set 
$P_4 = a_4\beta$. We obtain the final estimate
$\eta = -0.39(1)$,
which is fully consistent with those of Refs.~\onlinecite{HPV-08,Janus-13}.

Finally, we estimate $z$ by requiring data to satisfy the 
general scaling form (\ref{eq:R_vs_tLmz}). We obtain 
             $z = 6.80(15)$,
where the error should be quite conservative. The scaling behavior of 
$R_\xi$ and $U_4$ is shown in Fig.~\ref{fig:U4RxivstL}. Scaling corrections
are clearly visible, but large $L$ data appear to fall onto a single 
universal curve as $L$ increases. The value we obtain is in agreement 
with the estimate of Refs.~\onlinecite{Janus-z,FM-15} at $T=1.1$.
Instead,
it is larger than those of Refs.~\onlinecite{NEY-03,PC-05,LPSY-15}.
However, note that in all these works no 
scaling corrections, crucial to control possible systematic errors,
were included in the analyses (they play a fundamental 
role in the derivation of our result). 

Let us now summarize our results. We have presented a new dynamic 
off-equilibrium method suitable for the determination of the critical
temperature and of the critical exponents. Such a method represents a 
significant improvement with respect to previous ones. In particular,
there is no need for $L$ to be large enough to avoid finite-size 
effects---thus,
a source of systematic errors is absent---nor does it require an {\em a 
priori} knowledge of the critical temperature. We have used the method to 
determine critical exponents and temperature for the $\pm J$ Ising model. 
With a relatively modest investment of computing time, thanks also to a very 
efficient GPU multispin code, we obtain results that have a comparable 
precision with that of the estimates of Ref.~\cite{Janus-13}, 
which are the most precise equilibrium estimates available today.
The method is completely general and can be applied to any pure or disordered
system.

\end{document}